# Modelling of epitaxial graphene functionalization

D. W. Boukhvalov

*Computational Materials Science Center, National Institute for Materials Science,*

*1-2-1 Sengen, Tsukuba, Ibaraki 305-0047, Japan*

*A new model for graphene, epitaxially grown on silicon carbide is proposed. Density functional theory modelling of epitaxial graphene functionalization by hydrogen, fluorine and phenyl groups has been performed with hydrogen and fluorine showing a high probability of cluster formation in high adatom concentration. It has also been shown that the clusterization of fluorine adatoms provides midgap states in formation due to significant flat distortion of graphene. The functionalization of epitaxial graphene using larger species (methyl and phenyl groups) renders cluster formation impossible, due to the steric effect and results in uniform coverage with the energy gap opening.*

E-mail: D.Bukhvalov@science.ru.nl

# 1. Introduction

In recent times graphene has been discussed as the material with the most potential for post-silicon electronics [1-3]. The main requirement for graphene to have applications in electronic devices is its conversion from semimetal to semiconductor [4-6]. The most promising route for the realization of this transformation is through chemical functionalization [7-16]. Predicted theoretically, [15, 16] and realized experimentally,[17] the 100% hydrogenation of free standing graphene provides a metal-insulator transition. The one-side hydrogenation of graphene on a silicon dioxide ($SiO_2$) substrate has also led to the energy gap opening. The key role of substrate induced ripples for graphene functionalization and its transformation to semiconductor was discussed in our previous work [18, 19]. Exploring the functionalization of epitaxially grown graphene (further EG) on a silicon carbide substrate is the next step in the exploration of the role of substrate in graphene functionalization.

EG is one of the most suitable types of graphene [20-23] for large scale industrial production and application (see also for review [24, 25]). Recent experiments on EG hydrogenation [26, 27] at low hydrogen concentration reported a large number of hydrogen pairs previously discussed in theory [16, 28]. In contrast with the theoretical prediction approximately equal numbers of pars in *ortho* (Fig. 1a) and *para* (Fig. 1b) positions in experiments were observed. The further increase of hydrogen concentration led to the formation of small hydrogen clusters (similar to that shown on Fig. 1c) on the epitaxial graphene surface [27] instead of uniformly distributed adatoms (see Fig. 1d-e) with the energy gap opening modeled in first principles calculations [28]. The transition of EG to the semiconducting state were achieved only using aryl groups for the functionalization [29-31]. These disagreements between experimental results and theoretical predictions suggest that a special structural model of EG, taking into account substrate effects, should be considered to develop the methodology.

## 2. Model and computational method

Density functional theory (DFT) calculations have been carried out with the same pseudopotential code SIESTA [32] and Troullier-Martins pseudopotentials [33] used for our previous modelling of functionalized graphene [7, 12, 16-18, 28]. All calculations are done within local density approximation (LDA) [34] which is suitable for the description of multilayer graphene systems [12]. Detailed discussion about ability of the LDA functionals for the calculation of atomic and electronic structure of weakly bonded systems reported in Ref. 35. For the modelling of EG a supercell containing 32 carbon atoms in each layer (see Fig. 2) was used, within periodic boundary conditions. The distance between equivalent atoms in the supercells is 4 lattice parameters (about 1 nm). Thus the size of the supercell guarantees the absence of any overlap between distorted areas around chemisorbed adatoms [7]. For the modelling of different levels of binding with the SiC substrate, the hydrogenation coverage of the layer was varied between zero and 25%. All calculations were carried out for an energy mesh cut off 360 Ry and k-point mesh 8×8×2 in the Mokhorst-Park scheme [35]. During the optimization, the electronic ground state was found self-consistently using norm-conserving pseudo-potentials for cores and a double-ζ plus polarization basis of localized orbitals for carbon and oxygen, and double-ζ basis for hydrogen. Optimization of the forces and total energies was performed with an accuracy of 0.04 eV/Å and 1 meV, respectively. When drawing the pictures of density of states, a smearing of 0.2 eV was used. All calculations have been performed in spin polarized mode.

Due to the mismatch between graphene and SiC (0001) surface lattice parameters modeling of the exact atomic structure of EG requires usage of a large supercell containing several hundreds of atoms [37-40]. These models are suitable for studying subtle effects in the electronic structure of EG but extremely computationally expensive for research in chemical functionalization. Previous experimental and theoretical studies (see [24, 40] and references therein) suggest the presence of a carbon buffer layer between the graphene monolayer and the SiC surface. In our previous work [28] we found that a graphene bilayer with 25% adatom coverage of one of layer could be a suitable model for the description of the substrate induced energy gap opening in EG [40]. A graphene bilayer with functionalization of one layer with coverage lower than 25% (see Fig. 2a, b) could be a suitable model for studying EG chemical functionalization. One of the layers imitates the buffer layer and will be further called the under-layer (UL) and the functionalized layer will be referred to as the top-layer (TL). For minimization of the computational cost hydrogen were used as the adatoms for UL functionalization. The electronic structure of the proposed model (Fig. 3a) is close to that earlier reported for epitaxial graphene [36-38] and experimental ARPES measurement [41]. The coexistence of the carbon atoms with *sp2* and *sp3* hybridization results unusual band crossing between M-Γ and K-M symmetry point similar to the obtained for the exact model [38].

The chemisorption energy is the standard value used for characterization of functionalization, however, the used approximation (LDA) leads to overestimation of the chemisorption energy of the hydrogen on graphene [16]. The current work will examine the effect of different configurations of chemisorbed species and the total energy of the chemisorbed species could be used. This value is defined as $E_{specie} = (E_{host + N\ groups} - E_{host})/N$, where $E_{host}$ and $E_{host + N\ groups}$ is the total energy per graphene supercell before and after functionalization with N chemical species.

For the modeling of EG functionalization four types of species were used previously; hydrogen for free standing graphene [16] and EG functionalization [26, 27], fluorine as the most suitable substitute for hydrogen in graphene functionalization [7, 42-45], phenyl groups used as the model species for aryl groups in EG functionalization [29-31], and methyl (-$CH_3$) groups which is intermediate between single atom species and big functional groups such as phenyl. The previous first principles calculations of adatoms on graphene [7, 16] suggest that the formation of pairs of adatoms sitting on different graphene sublattices is much more energetically favorable (0.5 eV more per chemisorbed species) than chemisorption of single adatoms. Experimental results for graphene on SiC [26, 27], on $SiO_2$ [46] and graphite [47] also argue that the chemisorption of single adatoms or species on grapheme is disfavored. In the present work only the functionalization of graphene by pairs or even numbers of chemical species are considered.

## 3. Results and discussions

To check the method, chemisorption onto a grapheme monolayer was analyzed. The hydrogen pairs being in the *ortho* position (Fig. 1b) is less energetically favorable than the *para* position (Fig. 1a) by approximately 20 meV/H. The most favorable situation for the hydrogenation is uniform functionalization with a 25% coverage (Fig. 1e). The clustering (Fig. 1c) formation is much less stable (480 meV/H) than the uniform coverage. A result which corresponds with the formation of metastable hydrogen clusters on graphite surface [47]. In contrast with hydrogen, the formation of fluorine pairs in the *ortho* position is more favorable (40 meV/F) than in the *para* position. It is caused by the higher binding energy between carbon and fluorine and a stronger graphene flat corrugation of fluorinated graphene [7]. In the case of the methyl and phenyl groups (see Fig. 2c, d) the steric effect dictates that the formation of pairs in the *para* position much more stable than in the *ortho* position (about 2eV/phenyl) and much the more stable configuration corresponds with a uniform coverage of 1/8 of the carbon atoms (see Fig. 1d) [12, 48]. For numerically description of the steric effect for phenyl groups caused by London dispersion forces the value of the dispersion energy had been calculated with using standard formula $E_{disp} \approx -3\alpha^2 I/8R^6$, where $\alpha$ is the dipole polarizability [49], I – ionization potential [50] of phenyl groups and R the distance between two groups. The values of the dispersion energy is 1.25 eV and 0.22 eV for the distances 1.42 Å (*ortho* position) and 2.83 Å (*para* position) respectively. The total difference between the energy of the phenyl groups in *ortho* and *para* positions is the sum of the dispersion energies and the energy difference caused by the graphene sheet distortion [16]. The value of charge transfer from each hydrogen adatom to graphene is 0.11 electrons and from the methyl and phenyl groups it is 0.09 and 0.07 electrons respectively, independent of the level of coverage. This charge is calculated as a decrease in the numbers of electrons on the chemisorbed species. The transferred charge is distributed over the σ-orbitals of graphene and does not shift the Fermi level. As an oxidative species, fluorine withdraws 0.07 electrons from graphene. Further calculations for the graphene bilayer with different levels of hydrogenation of UL suggest invariance in the number of transferred electrons for all types of studied graphenes. The charge transfer from the hydrogen atoms chemisorbed onto the UL is also 0.11 electrons and is distributed amongst the σ-orbitals of the UL.

In the case of a graphene bilayer with a non-functionalized UL, the chemisorption scenarios are the same as for the graphene monolayer. But hydrogenation of the UL dramatically changes the scenarios of the TL functionalization. Independent from the level of UL coverage for hydrogens, on the TL the energy difference between pair formation in *ortho* and *para* position is negligible (less than 10 meV/H), meaning the formation of both types of pairs are probable, and that has been experimentally observed by two groups [26, 27]. These hydrogen pairs will act as the centers for the further clusterization (Fig. 1c). The energy difference between clustering and uniform hydrogenation is also rather small (about 10 meV/H; independent from the level of UL functionalization). To check the validity of the model calculations, the larger supercell contained 72 carbon atoms in each layer, and calculations have been performed for all hydrogen pairs and clusters. Deviations from the results obtained for by the smaller model were less that 2 meV/H. In contrast with corrugated graphene with midgap states [18] the described clusterization of hydrogen does not provide the energy gap opening (see Fig. 4). The cause of this dramatic difference between the cases of non-hydrogenated and hydrogenated UL is changes in the weak interactions between layers due to both of their functionalization. Prior to functionalization of the graphene bilayer, two sublattices in each layer can be denoted. One of these sublattices connects by π-π bond with the neighboring layer. Due to the hydrogenation of several carbon atoms, each sublattice is divided into two new sublattices that violate the homogeneity of interlayer binding and lead to changes in the energy differences between different configurations of chemisorbed adatoms. In the fluorination scenario however, after functionalization the UL remains the same. The energy difference between pairs in the *ortho* and *para* positions increases from 45 meV/F for the pure bilayer to 62 meV/F in the case of the hydrogenated UL. The crucial difference between the formation of hydrogen and fluorine clusters is the bigger distortion of the TL from the flat (0.97 and 0.64 Å respectively). This enormous distortion of the graphene sheet provides the formation of midgap states at Fermi level (see Fig. 4c) similarly to the case of the corrugated graphene monolayer [18].

Similarly to the above case of single layer graphene functionalization, using methyl and phenyl species, the *para* position is much more favorable than *ortho* due to the aforementioned steric effect (see Fig. 2c, d), making clusterization impossible. Similarly to the pure graphene bilayer, uniform functionalization with coverage of 12.5% is the most stable. The hydrogenation of the graphene UL causes weakening of the interlayer bonds [12] and the decay of the total energy caused by these groups is about 18 meV/methyl and 24 meV/phenyl compared to a non-functionalized graphene bilayer. The obtained value of coverage for the most stable configuration is near to the experimentally observed ratio of 10:1 of *sp2* and *sp3* picks in the XPS core level spectra of aryl-functionalized EG [29]. The electronic structure calculated for the studied configuration gives evidence for an energy gap (18 mev for methyl and 15 meV for phenyl) opening (Fig. 3b). The calculated value of the energy gap obtained for the level of coverage of both layers is lower than in our previous work [28] and in agreement with the experimental transition of aryl-functionalized EG from semimetal to semiconductor [29-31] and the shape of the bands near Fermi level is also similar to the experimentally measured spectras [31]. For further validate used model and explore the role of substrate in energy gap opening, calculations of the same structure with UL fluorination instead of hydrogenation were performed. The energy gap value increased insignificantly from 15 to 18 meV for the graphene functionalized by the phenyl groups.

## 4. Conclusions

The proposed model of the epitaxial graphene as graphene monolayer with partial single-layer functionalization provide an explanation of the experimentally observed (i) presence of *para* and *ortho* pairs of hydrogen, (ii) formation of stable hydrogen clusters for a higher degree of functionalization, (iii) energy gap opening in the case of aryl-functionalization. The observed coexistence of a small energy gap and a uniform distribution when epitaxial graphene is functionalized with various large chemical species makes this material very promising for applications in nanoelectronic devices. The developed model could be used for further modeling of epitaxial graphene functionalization, taking into account the GW-methods correction to the gap values calculated within LDA methods [10].

**Acknowledgements** I gratefully acknowledge G. N. Newton for careful reading the manuscript.

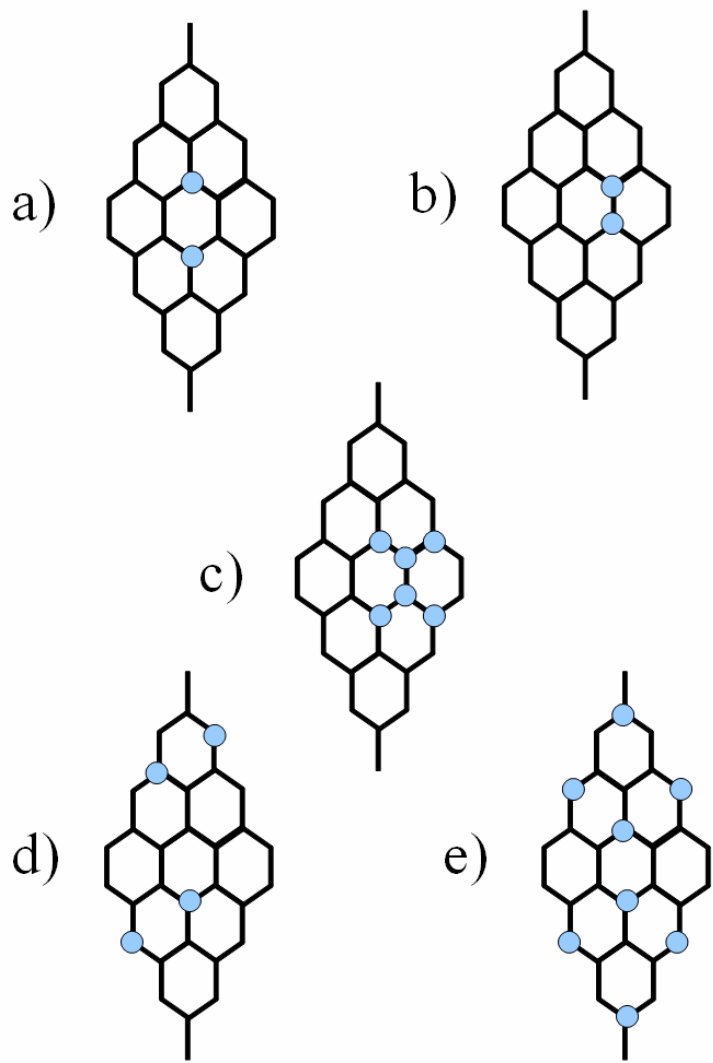

**Fig. 1** A sketch of the different scenarios of one-side graphene functionalization. Formation of pair of adatoms in *para* (a) and *ortho* (b) position, cluster of six adatoms (c), and uniform coverage of 12.5% (d) and 25% (e) of graphene surface.

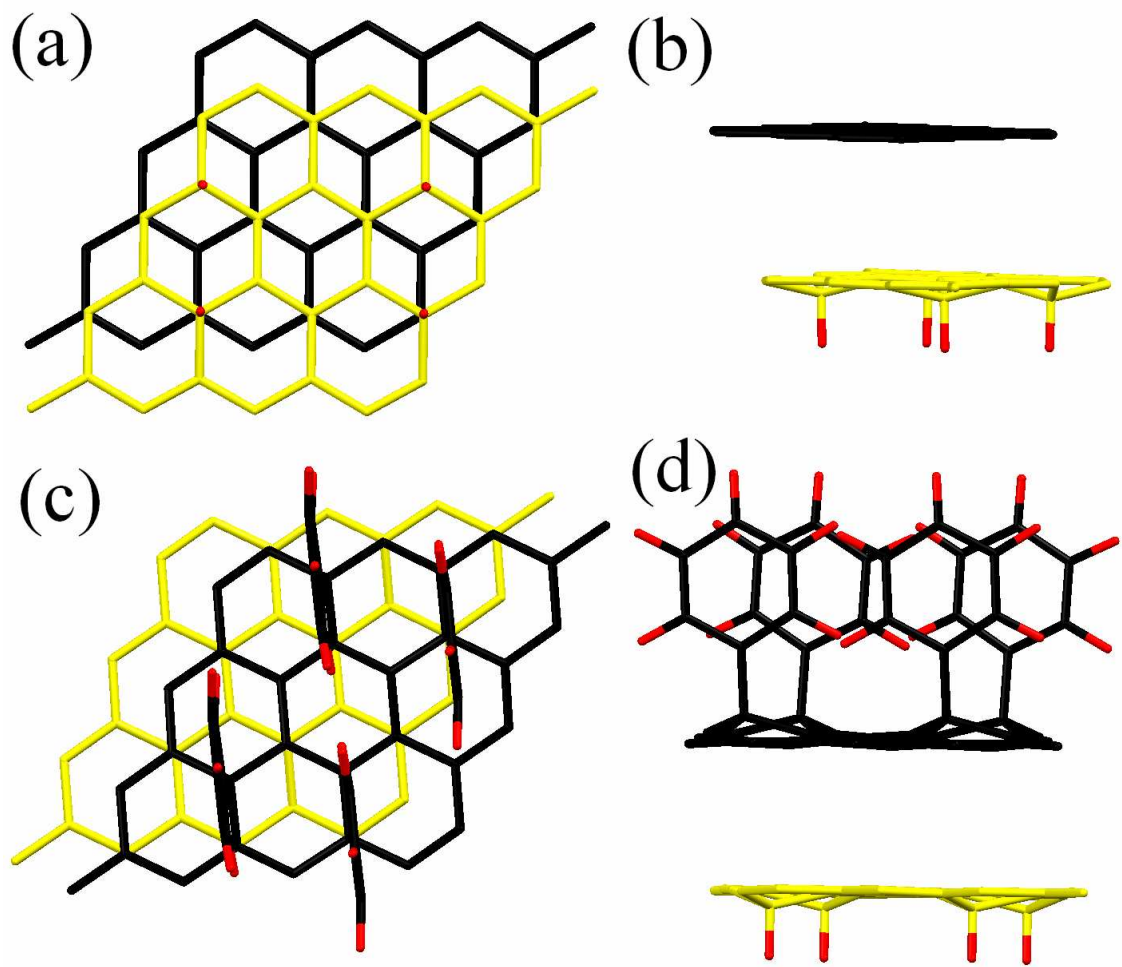

**Fig. 2** Bottom (a) and side (b) view of optimized atomic structure of graphene bilayer with 12.5% hydrogenated UL, and top (c) and side (d) view of optimized atomic structure of graphene bilayer with 15.5% coverage of TL by phenyl groups and same coverage of UL by hydrogen. Carbon atoms of TL and phenyl groups are shown by black, carbon atoms of UL by yellow, and hydrogen atoms by red.

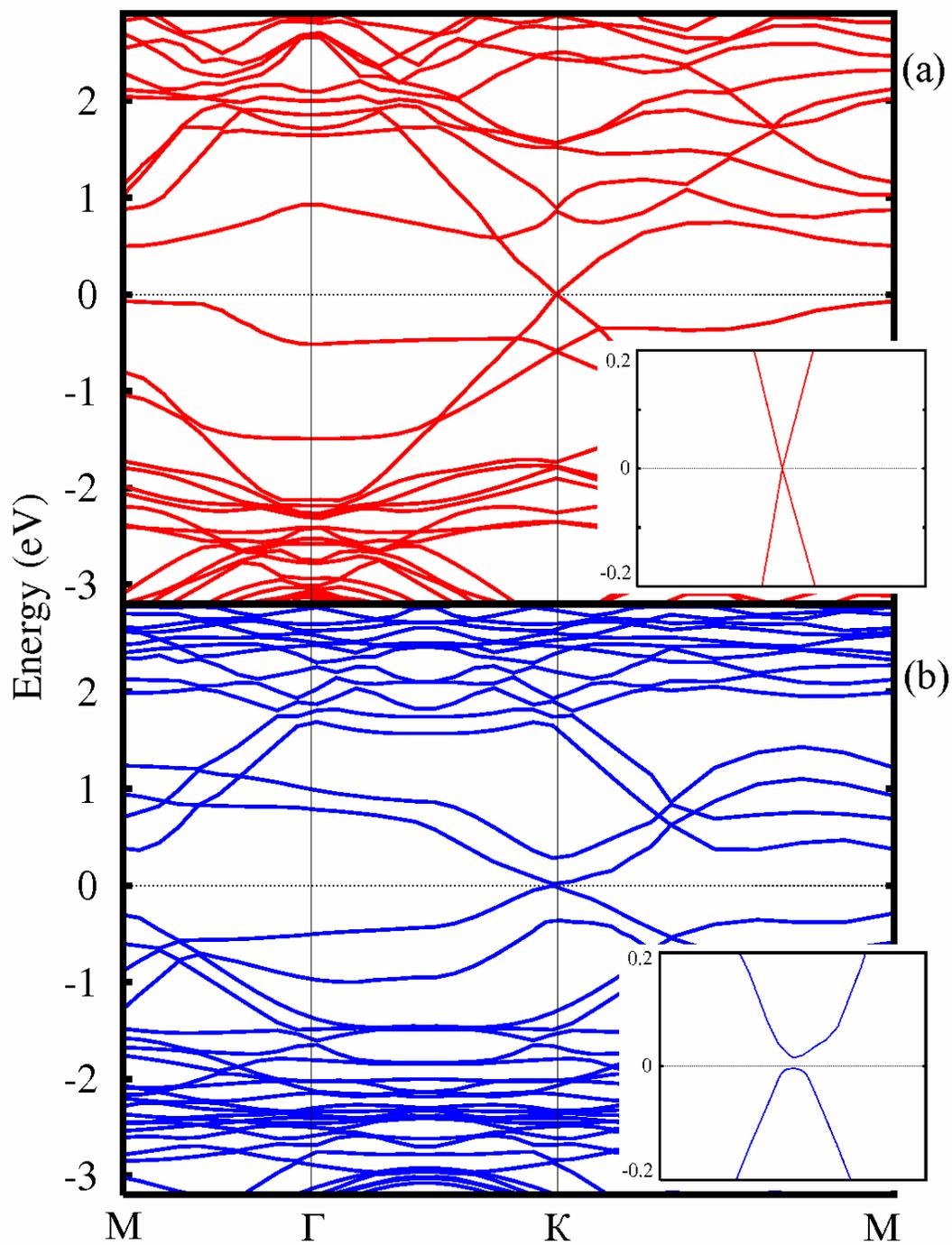

**Fig. 3** Band structure of (a) graphene bilayer with non-functionalized TL and hydrogenated with 12.5% coverage UL (see Fig. 2a, b), and (b) graphene bilayer with 15.5% coverage of TL by phenyl groups and same coverage of UL by hydrogen (see Fig. 2c, d). On insets higher resolution band structures near the K point.

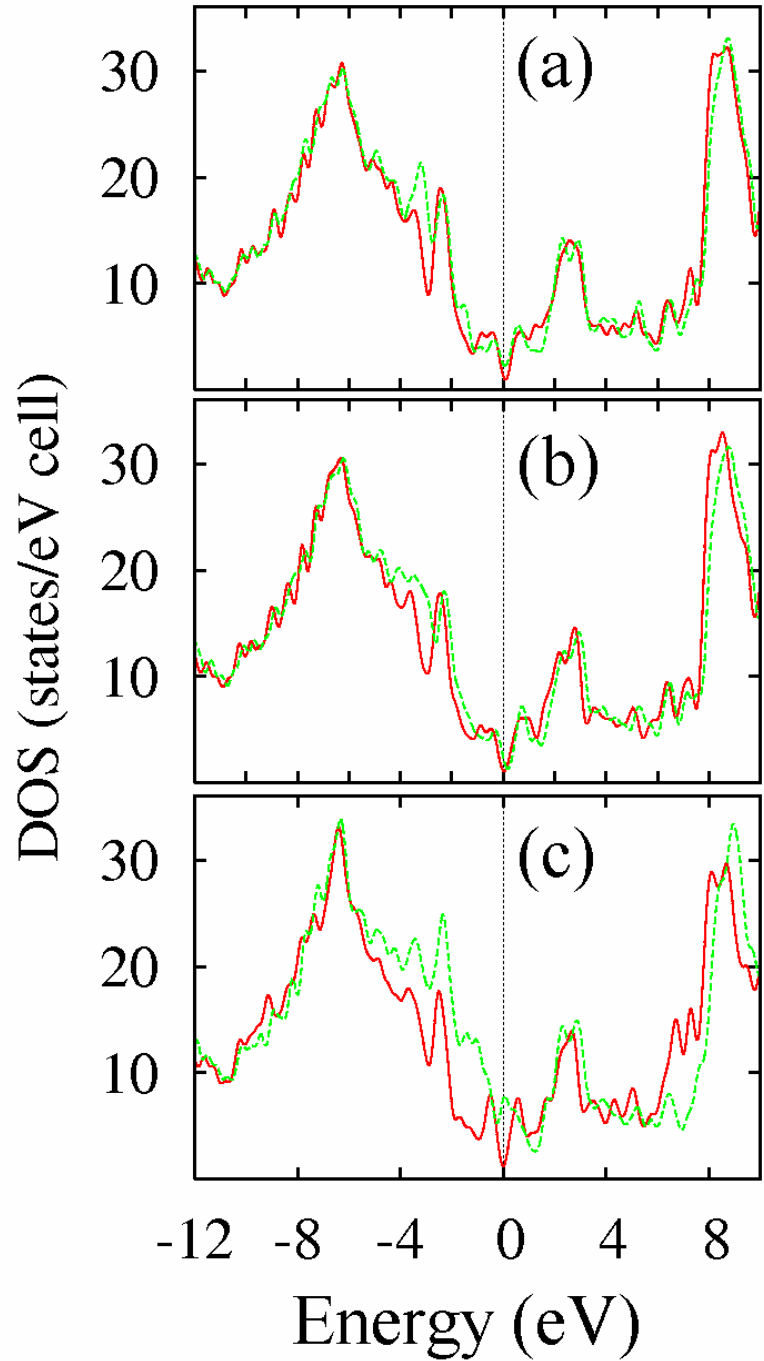

**Figure 4** Densities of states for graphene bilayer with 12.5% hydrogenated UL and TL functionalized by hydrogen (solid red line) or fluorine (dashed green line) with formation of the pairs in *para* (a) and *ortho* (b) positions and also cluster of six adatoms (c). All discussed structures are equivalent to those presented in Fig. 1a-c.